\documentstyle [12pt,epsf]{article}

\input{epsf}
\begin{document}
\begin{titlepage}
\begin{center}

\vspace{-0.8in}

{\large \bf Path integral approach to the full Dicke model\\ with  dipole-dipole interaction}\\
\vspace{.2in}{\large\em M. Aparicio Alcalde\,$^{a,}$\footnotemark[1], J. Stephany\,$^{b,c,}$\footnotemark[2],
and N. F. Svaiter\,$^{d,}$\footnotemark[3]}\\
\vskip.1in

$~^{a}$
Instituto de F\'{\i}sica Te\'orica, UNESP - S\~ao Paulo State University,
Caixa Postal 70532-2, 01156-970 S\~ao Paulo, SP, Brazil.

\vskip.1in
$~^{b}$
Departamento de F\'{\i}sica, Secci\'{o}n de Fen\'{o}menos \'{O}pticos, \\
Universidad Sim\'{o}n Bol\'{\i}var, Apartado Postal 89000, Caracas 1080-A, Venezuela.\\

\vskip.1in
$~^{c}$
Max-Planck-Institut f\"ur Gravitationsphysik, Albert-Einstein-Institut\\
Am M\"uhlenberg 1, 14476 Golm, Germany.\\

\vskip.1in
$~^{d}$
Centro Brasileiro de Pesquisas F\'{\i}sicas, Rua Dr. Xavier Sigaud 150,\\
22290-180, Rio de Janeiro, RJ Brazil.\\

\subsection*{\\Abstract}
\end{center}

\baselineskip .18in

We consider the full Dicke spin-boson model composed by a single
bosonic mode and an ensemble of $N$ identical two-level atoms with different
couplings for the resonant
and anti-resonant interaction terms, and incorporate a dipole-dipole interaction between the atoms.
Assuming that the system is in thermal equilibrium with a
reservoir at temperature $\beta^{-1}$, we compute the free energy in the thermodynamic limit 
$N\rightarrow\infty$ in the
saddle-point approximation to the path integral and determine the critical
temperature for the superradiant phase transition. In the zero temperature limit,
we recover the critical coupling of the quantum phase transition, presented in the
literature.

\vskip.1in
PACS numbers: 42.50.Fx, 05.30.Jp, 73.43.Nq

\footnotetext[1]{e-mail:\,\,aparicio@ift.unesp.br}
\footnotetext[2]{e-mail:\,\,stephany@usb.ve}
\footnotetext[3]{e-mail:\,\,nfuxsvai@cbpf.br}

\end{titlepage}
\newpage\baselineskip .18in
\section{Introduction}
\quad\quad $\,\,$

At low temperatures a gas of N slow moving  molecules with a few accessible electronic states
interacting with the vacuum electromagnetic field in a cavity may be considered  as a single
quantum mechanical system with a ground state which determines the macroscopic properties of
the gas and a set of excited states emerging as collective  matter-field modes.  In
conditions where the dependence on the center of mass coordinates of the molecules may
be omitted and for the simplest case  of two level atoms, the system may be
approximately described by Dicke \cite{dicke} and generalized Dicke models which
are written in terms of equivalent spin variables for the atoms.
At zero temperature the properties of these models can be investigated by solving
the quantum mechanical problem in a suitable approximation. In the original Dicke
model \cite{dicke}  it is furthermore assumed that the atoms are confined to a small
volume and that their direct interaction may be disregarded. The field is coupled
to the total equivalent  spin operator and enhanced rates for spontaneous radiation
may be computed in the dipolar approximation for  particular, so called superradiant
states. In a modern view this is recognized  as a hint of the existence of  a quantum phase
transition for this system \cite{sachdev}. A direct  determination of this transition
was obtained in Ref. \cite{emary} performing the diagonalization of the bimodal effective Hamiltonian resulting
of applying a Holstein-Primakoff transformation \cite{holstein,lawande} in the thermodynamic
limit and computing the critical coupling for which the boson expectation value vanish.
In this work the energy of the excited collective modes was also computed and the relation of the quantum phase transition and the entanglement properties of the collective states was discussed. This connection is particularly interesting due to the wide interest that have attracted the applications to communications and quantum computing of entangled systems  \cite{livro}. In this context  efforts are focused in how to create and also how to detect entangled states in many-body systems \cite{cm2,cm3,amico}, but many authors have also investigated the relation between entanglement of collective modes and quantum phase transitions  \cite{osborn,amico1,kitaev,lambert}.

An alternative approach to the study of this model, which in fact was pursued much earlier is
to explore the thermodynamical properties of the system at finite temperature by computing the
partition function and the free energy. This was done first in the rotating wave approximation
by Hepp and Lieb \cite{hepp,hepp2} and Wang and Hioe \cite{wang} who determined the critical
temperature for the superradiant phase transition in the Dicke model. In Ref. \cite{hioe} a generalized
Dicke model which include both the resonant and anti-resonant interaction terms was considered . The results in this works were confirmed in \cite{vertogen,pimentel} using different methods.

The use of the path integral approach for the computation of the thermodynamical properties of
these models was advocated by Popov and their collaborators \cite{popov1,popov2,popov3,popov4, popov5}
and produce results which are consistent with those obtained by other techniques in a very efficient way.
The path integral formalism provides in particular a simple method to compute the spectrum of the system
under consideration in either of it phases using the semiclassical expansion of the partition
function \cite{popov3,yarunin}.  In Refs. \cite{aparicio1,aparicio2} these methods where used
to determine the partition function and critical temperature  of the generalized  Dicke model and
of the full Dicke model which emerges when different couplings for the resonant and anti-resonant
terms are introduced. Then in Ref. \cite{aparicio3} the spectrum of these models was determined by
computing the quadratic approximation of the path integral. The zero temperature limit of the
energies, in the case when the resonant and anti-resonant couplings are the same, were shown to
correspond to the result in \cite{emary}.

When  dipole-dipole  interactions are included the procedure using the Holstein Primakoff
transformation is still useful and the critical coupling corresponding to the zero temperature quantum
phase transition as well as the energy of the excited states may be computed \cite{chen,nie}.
See also Refs. \cite{chen1,pan,lee}. 

In this paper we compute the free energy in this generalized situation using functional methods and 
determine the critical temperature.   Our  result show a
dependence of the critical coupling on the dipole-dipole coupling which in the zero temperature limit 
reproduce consistently the quantum phase
transition point computed in  \cite{chen}.  The organization of the paper is as follows:
In section II we discuss the interacting Hamiltonian of the model. In section III the functional integral
for the model is performed and the critical temperature is determined. In section IV we present the free energy of the model.
Finally, section V contains our conclusions. We use $k_{B}=c=\hbar=1$

\section{The Hamiltonian and the fermionic model}

We consider the full Dicke model with dipole-dipole interaction. This model contemplates resonant and
anti-resonant coupling terms between the atoms and the bosonic field with different couplings  $g_1$
and $g_2$. The Hamiltonian for this system, with the dipole-dipole interaction included reads
\begin{eqnarray}
 H&=& \frac{\lambda}{N}\sum_{i\neq
j}^{N}\sigma_{i}^{+}\,\sigma_{j}^{-} +\,\frac{\Omega}{2}\,\sum_{i=1}^{N}
\,\sigma_{i}^{z}+\omega_0\,b^{\dagger}\,b\,\nonumber\\
&+&
\frac{g_1}{\sqrt{N}}\sum_{i=1}^{N} \left(
b\,\sigma_{i}^{+}+b^{\dagger}\,\sigma_{i}^{-}\right)\,+
\frac{g_2}{\sqrt{N}}\sum_{i=1}^{N} \left(
b\,\sigma_{i}^{-}+b^{\dagger}\,\sigma_{i}^{+}\right).
\label{Hdipole}
\end{eqnarray}
In the above equation  the operators $\sigma_{i}^1$,
$\sigma_{i}^2$ and $\sigma_{i}^z=\sigma_{i}^3$ satisfy the commutation
relations $[\sigma_{i}^p,\sigma_{i}^q]=
2\,\epsilon^{pqr}\,\sigma_{i}^r$ for $p,q,r=1,2,3$ and we use through the paper the 
operators 
$\sigma_{i}^{\pm}=\frac{1}{2}\,(\sigma_{i}^{1} \pm i\,\sigma_{i}^{2})$. 
These operators satisfy $[\sigma_{i}^+,\sigma_{i}^-]=\sigma_{(j)}^z$
and $[\sigma_{i}^z,\sigma_{i}^{\pm}]=\pm\,2\,\sigma_{i}^{\pm}$. The
boson annihilation and creation
operators $b$ and $b^{\dagger}$  satisfy the usual commutation
relation rules.  Each two-level atom interact with all other atoms of the ensemble
with the same coupling strength $\lambda/N$, and the
summation is over all the atoms. Note that this model is related to models describing
spin systems with long range interaction \cite{haldane,shastry,pittel}.

To proceed, let us define an auxiliary model to be called the fermionic full Dicke model in terms of
fermionic raising and lowering operators
$\alpha^{\dagger}_{i}$, $\alpha_{i}$, $\beta^{\dagger}_{i}$ and
$\beta_{i}$, that satisfy the anti-commutator relations
$\alpha_{i}\alpha^{\dagger}_{j}+\alpha^{\dagger}_{j}\alpha_{i}
=\delta_{ij}$ and
$\beta_{i}\beta^{\dagger}_{j}+\beta^{\dagger}_{j}\beta_{i}
=\delta_{ij}$. We can also define the following bilinear
combination of fermionic operators, $\alpha^{\dagger}_{i}\alpha_{i}
-\beta^{\dagger}_{i}\beta_{i}$, $\alpha^{\dagger}_{i}\beta_{i}$
and $\beta^{\dagger}_{i}\alpha_{i}$ which obey
the same commutation relations as the pseudo-spin operators
$\sigma^z_{(\,i)}$, $\sigma^+_{(\,i)}$ and $\sigma^-_{(\,i)}$ with the correspondence given by,
\begin{equation}
\sigma_{i}^{z}\longrightarrow \alpha_{i}^{\dagger}\alpha_{i}
-\beta_{i}^{\dagger}\beta_{i}\, , \label{34}
\end{equation}
\begin{equation}
\sigma_{i}^{+}\longrightarrow \alpha_{i}^{\dagger}\beta_{i}\, ,
\label{35}
\end{equation}
\begin{equation}
\sigma_{i}^{-}\longrightarrow \beta_{i}^{\dagger}\alpha_{i}\, .
\label{36}
\end{equation}
Substituting these relations in the Hamiltonian (\ref{Hdipole}) we obtain the Hamiltonian $ H_F$ of the 
auxiliary fermionic full Dicke model with dipole-dipole coupling,
\begin{eqnarray}
&& H_F= \frac{\lambda}{N}\sum_{ij}^{N}\alpha_{i}^{\dagger}\beta_{i}\,\beta_{j}^{\dagger}\alpha_{j}
+\,\frac{\Omega}{2}\,\sum_{i=1}^{N}
\,(\alpha_{i}^{\dagger}\alpha_{i}
-\beta_{i}^{\dagger}\beta_{i}) \nonumber\\
&&+\,\omega_0\,b^{\dagger}\,b\,+
\frac{g_1}{\sqrt{N}}\sum_{i=1}^{N} \left(
b\,\alpha_{i}^{\dagger}\beta_{i}+b^{\dagger}\,\beta_{i}^{\dagger}\alpha_{i}\right)\,+
\frac{g_2}{\sqrt{N}}\sum_{i=1}^{N} \left(
b\,\beta_{i}^{\dagger}\alpha_{i}+b^{\dagger}\,\alpha_{i}^{\dagger}\beta_{i}\right).
\label{HFdipole}
\end{eqnarray}
The Hamiltonians $H$ and $H_F$ are defined in different Hilbert spaces, since each operator
$\sigma_i^{\alpha}$ in $H$ acts on a two-dimensional Hilbert space,
and each of the fermionic operators $\alpha^{\dagger}_{i}$, $\alpha_{i}$,
$\beta^{\dagger}_{i}$ and $\beta_{i}$, in
$H_F$ acts on a four-dimensional Fock space. Nevertheless the partition
function $Z$ of the Dicke model may be written as a trace in the space of the fermionic
operators due to the the following useful relation whose demonstration follows the one 
given by Popov for the Dicke model \cite{popov2}
\begin{eqnarray}
Z=Tr\Bigl(\exp(-\beta\,H)\Bigr)=i^N\,Tr\left(\exp\left(-\beta\,H_F-\frac{i\pi}{2}\,N_F\right)\right)\,.
\label{partitionsfunctions}
\end{eqnarray}
In this relation $H$ is given by  Eq. (\ref{Hdipole}), $H_F$ is given by Eq. (\ref{HFdipole})
and the operator $N_F$ is defined by
\begin{eqnarray}
N_F=\sum_{i=1}^{N}(\alpha_{i}^{\dagger}\alpha_{i}+\beta_{i}^{\dagger}\beta_{i})\,.
\end{eqnarray}
It should be  clear that the traces  in Eq. (\ref{partitionsfunctions}) for each Hamiltonian are taken over their respective spaces.

\section{The partition function in the path integral approach}

As mentioned in the introduction the critical temperature for the Dicke and the full Dicke models may be computed using functional methods \cite{aparicio2,aparicio3}. In this section we consider the corresponding computation for the full Dicke model with dipole-dipole interaction. This problem was addressed earlier in Ref. \cite{nami}, but within the approximation scheme proposed there no variation of the critical temperature was detected. In this section we prove that  the introduction of the dipole-dipole interaction indeed modify the temperature for transition from normal to superradiant phase. Concerning the 
path integral approach that we use, we note that in this case due to the dipole-dipole
interaction term the path integral is non Gaussian. To overcome this difficulty  we introduce an auxiliary field in the path integral and perform a series expansion..

At finite temperature the Euclidean action $S$ associated to the fermionic Dicke model is given by
\begin{equation}
S=\int_0^{\beta} d\tau \left(b^*(\tau)\,\partial_{\tau}b(\tau)+ \sum_{i=1}^{N}
\Bigl(\alpha^*_i(\tau)\,\partial_{\tau}\alpha_i(\tau)
+\beta^*_i (\tau)\,\partial_{\tau}\beta_i(\tau)\Bigr)\right) -\int_0^{\beta}d\tau H_{F}(\tau)\,,
\label{66}
\end{equation}
where the Hamiltonian density $H_{F}(\tau)$ is obtained from Eq. (\ref{HFdipole}). In this case it takes the form
\begin{eqnarray}
H_{F}(\tau)&=&\omega_{0}\,b^{\,*}(\tau)\,b(\tau)\,+
\,\frac{\Omega}{2}\,\displaystyle\sum_{i\,=\,1}^{N}\,
\biggl(\alpha^{\,*}_{\,i}(\tau)\,\alpha_{\,i}(\tau)\,-
\,\beta^{\,*}_{\,i}(\tau)\beta_{\,i}(\tau)\biggr)\,+
\nonumber\\
&+&\frac{\lambda}{N}
\sum_{i\neq
j}^{N}\,\alpha^{\,*}_{\,i}(\tau)\,\beta_{\,i}(\tau)\,
\beta^{\,*}_{\,j}(\tau)\alpha_{\,j}(\tau)+\frac{g_{\,1}}{\sqrt{N}}
\,\displaystyle\sum_{i\,=\,1}^{N}\,
\biggl(\alpha^{\,*}_{\,i}(\tau)\,\beta_{\,i}(\tau)\,b(\tau)\,+
\beta^{\,*}_{\,i}(\tau)\,\alpha_{\,i}(\tau)\,b^{\,*}(\tau)\,\biggr)\,+
\nonumber\\
&+&\frac{g_{\,2}}{\sqrt{N}}\,\displaystyle\sum_{i\,=\,1}^{N}\,
\biggl(\beta^{\,*}_{\,i}(\tau)\,\alpha_{\,i}(\tau)\,b(\tau)\,+
\,\alpha^{\,*}_{\,i}(\tau)\,\beta_{\,i}(\tau)\,b^{\,*}(\tau)\biggr).
\label{b32}
\end{eqnarray}
Now we consider  the formal ratio of the partition function of the dipole-dipole full Dicke
model and the partition function of the free Dicke model,
\begin{equation}
\frac{Z}{Z_0}=\frac{\int [d\eta(\alpha,\beta,b)]\,\exp{\left(\,S-\frac{i\pi}{2\beta}\int_0^{\beta}n(\tau)d\tau\right)}}{\int
[d\eta(\alpha,\beta,b)]\,\exp{\left(\,S_{0}-\frac{i\pi}{2\beta}\int_0^{\beta}n(\tau)d\tau\right)}}\, ,
\label{65}
\end{equation}
where the function $n(\tau)$ in the above expression is defined by
\begin{eqnarray}
n(\tau)=\sum_{i=1}^{N}
\Bigl(\alpha^{\,*}_i(\tau)\,\alpha_i(\tau)+\beta^{\,*}_i(\tau)\beta_i(\tau)\Bigr)\,.
\end{eqnarray}
In  Eq. (\ref{65}) $S=S(b,b^*,\alpha,\alpha^{\dagger},\beta,\beta^{\dagger})$
is the Euclidean action of the full Dicke model
given by Eq. (\ref{66}),
$S_0=S_{0}(b,b^*,\alpha,\alpha^{\dagger},\beta,\beta^{\dagger})$
is the free Euclidean action obtained by
taking $\lambda=g_1=g_2=0$ and  $[d\eta(\alpha,\beta,b)]$ is the functional
measure.
The functional integrals  in Eq. (\ref{65}), have to be done  in the space of complex functions
$b^*(\tau)$ and $b(\tau)$ and Grassmann variables
$\alpha_i^*(\tau)$, $\alpha_i(\tau)$, $\beta_i^*(\tau)$ and
$\beta_i(\tau)$. In thermal equilibrium in the imaginary time formalism, the integration
variables in Eq. (\ref{65}) obey periodic boundary conditions for
the Bose field, i. e., $b(\beta)=b(0)$ and anti-periodic boundary
conditions for Grassmann variables, i. e., $\alpha_i(\beta)=-\alpha_i(0)$
and $ \beta_i(\beta)=-\beta_i(0)$.

To advance in the computation of the quotient (\ref{65}), we perform now the following decoupling transformation
\begin{eqnarray}
\begin{array}{cc}
\alpha_i(\tau)\rightarrow e^{\frac{i\pi}{2\beta}\tau}\,\alpha_i(\tau)\,,\;\;\;  &
\alpha_i^*(\tau)\rightarrow e^{-\,\frac{i\pi}{2\beta}\tau}\,\alpha_i^*(\tau)\,,\\
\beta_i(\tau)\rightarrow e^{\frac{i\pi}{2\beta}\tau}\,\beta_i(\tau)\,,\;\;\;  &
\beta_i^*(\tau)\rightarrow e^{-\,\frac{i\pi}{2\beta}\tau}\,\beta_i^*(\tau)\,.
\end{array}
\label{trans2}
\end{eqnarray}
which makes the coefficient of the term proportional to  $n(\tau)$ in the action to vanish.
We should  remark that the price to pay for this benefit is that the boundary conditions are modified. In
Eq. (\ref{65n}), the Bose field still obeys periodic boundary conditions, i. e., $b(\beta)=b(0)$,
but the Fermi fields obey now the following boundary conditions:
\begin{eqnarray}
\begin{array}{cc}
\alpha_i(\beta)=i\,\alpha_i(0)\,,\;\;\;&
\alpha_i^*(\beta)=-\,i\,\alpha_i^*(0)\,,\\
\beta_i(\beta)=i\,\beta_i(0)\,,\;\;\;&
\beta_i^*(\beta)=-\,i\,\beta_i^*(0)\,.
\end{array}
\label{nbound}
\end{eqnarray}
After this procedure we obtain
\begin{equation}
\frac{Z}{Z_0}=\frac{\int [d\eta(\alpha,\beta,b)]\,e^S}{\int[d\eta(\alpha,\beta,b)]\,e^{S_0}}\,.
\label{65n}
\end{equation}
In order to obtain the effective action of the bosonic mode we
must integrate over the Grassmann Fermi fields in  Eq. (\ref{65n}). The problem that we now
are confronting
is  that in the action given by Eq. (\ref{b32}) there is a quartic term corresponding to the
dipole-dipole interaction and the integral is non Gaussian.
Although it is not possible to integrate this term directly, we can use an
auxiliary field to circumvent this
difficult and express the path integral as a series expansion. Since  terms quadratic in the
Grassman variables obey Bose statistics  we have
$(\alpha_i(\tau)\,\beta^*_i(\tau))\,(\alpha_j(\tau)\,\beta^*_j(\tau))
=(\alpha_j(\tau)\,\beta^*_j(\tau))\,(\alpha_i(\tau)\,\beta^*_i(\tau))$,
and we can use following functional identity
\begin{eqnarray}
&&e^{-\frac{\lambda}{N}\sum_{i,j=1}^N \alpha^*_i(\tau)\beta_i(\tau)\,
\beta^*_j(\tau)\alpha_j(\tau)}\nonumber\\
&=&N_0\,\int [d\eta(r)]\,
e^{\int_0^{\beta}d\tau\left(r^*(\tau)\,r(\tau)\,+\,\sqrt{\frac{\lambda}{N}}
\sum_{i=1}^N r(\tau)\,\beta^*_i(\tau)\alpha_i(\tau)\,+\,
\sqrt{\frac{\lambda}{N}}\sum_{i=1}^N
r^*(\tau)\,\alpha^*_i(\tau)\beta_i(\tau)\right)}\,,
\label{intrealfermion}
\end{eqnarray}
where $[d\eta(r)]$ is the functional measure for the functional integral with respect
the fields $r(\tau)$ and $r^*(\tau)$. The normalization factor is defined by $N_0^{-1}=\int [d\eta(r)]\,\exp{\Bigl(\int_0^{\beta}d\tau\,r^*(\tau)\,r(\tau)\Bigr)}$,
and the fields $r(\tau)$ and $r^*(\tau)$ satisfy the  boundary conditions
$r(0)=r(\beta)$ and $r^*(0)=r^*(\beta)$.
Substituting this last expression in  $Z$,
we obtain
\begin{eqnarray}
Z=N_0\,\int [d\eta(r)]\,[d\eta(\alpha,\beta,b)]\,e^{\,S_r}\,.
\label{part1}
\end{eqnarray}
The new action $S_r$ can be separated into a free action $S_{0}(b,r)$ 
for the bosons and the field $r(\tau)$ and a Gaussian fermionic part in the form,
\begin{equation}
\label{Sr}
S_r = S_{0}(b,r) +  \int_{0}^{\beta} d\tau\,\sum_{i=1}^{N}\,
\rho^{\dagger}_{i}(\tau)\, M_r(b^{*},b)\,\rho_{i}(\tau)\, ,
\label{pseudoacao}
\end{equation}
with
\begin{equation}
S_{0}(b,r) = \int_{0}^{\beta} d\tau\,\left( b^{*}(\tau)\,\Bigl(
\partial_{\tau}-\omega_{0}\Bigr)\,b(\tau)\,+\,r^*(\tau)\,r(\tau)\right)\,,
\label{67}
\end{equation}
In (\ref{Sr}), the column matrix $\rho_{\,i}(\tau)$ arranges the
Grassmann Fermi fields in the form
\begin{eqnarray}
\rho_{\,i}(\tau) &=& \left(
\begin{array}{c}
\beta_{\,i}(\tau) \\
\alpha_{\,i}(\tau)
\end{array}
\right),
\nonumber\\
\rho^{\dagger}_{\,i}(\tau) &=& \left(
\begin{array}{cc}
\beta^{*}_{\,i}(\tau) & \alpha^{*}_{\,i}(\tau)
\end{array}
\right) \label{69a}
\end{eqnarray}
and the matrix $M_r(b^{*},b)$ is given in terms of the operators $L=\partial_{\tau} + \Omega/2$ and $L_*=\partial_{\tau} - \Omega/2$ by
\begin{eqnarray}
\left( \begin{array}{cc}
L &  -N^{-1/2}\,\biggl(g_{1}
\,b^{*}\,(\tau) + g_{2}\,b\,(\tau)-\sqrt{\lambda}\,r(\tau)\biggr)\\
-N^{-1/2}\,\biggl(g_{1}\,b\,(\tau) + g_{2}\,b^{*}\,(\tau)-\sqrt{\lambda}\,r^*(\tau)\biggr)
 & L_*
\end{array} \right)\,,
\label{matrix1}
\end{eqnarray}
From the form of $S_r$ we see that the fermionic part of the functional integral given by Eq. (\ref{part1}) is
Gaussian and can be done directly. Therefore we obtain
\begin{eqnarray}
Z=N_0\,\int [d\eta(r)]\,[d\eta(b)]\,e^{S_{0}(b,r)} \Bigl(\det{M_r(b^{*},b)}\Bigr)^N\,,
\label{Zop}
\end{eqnarray}
where in this case, $[d\eta(b)]$ is the functional measure only for the bosonic field.
With the help of the following property for matrices with operator components
\begin{eqnarray}
\det\left(\begin{array}{cc}
A&B\\
C&D
\end{array}
\right)=\det\left(AD-ACA^{-1}B\right)\,,
\label{mazprop}
\end{eqnarray}
and using also determinant properties, we finally obtain,
\begin{eqnarray}
&&\det{M_r(b^{*},b)}=\nonumber\\
&&\det{\Bigl(LL_*\Bigr)}\,\det{\left(1-N^{-1}L_*^{-1}
\Bigl(g_1\,b+ g_2\,b^*-\sqrt{\lambda}\,r^*\Bigr)L^{-1}\Bigl(
g_1\,b^*+ g_2\,b-\sqrt{\lambda}\,r\Bigr)\right)}\,.
\label{Mop1}
\end{eqnarray}
Substituting Eq. (\ref{Zop}) and Eq. (\ref{Mop1}) in Eq. (\ref{65n}), we have
\begin{eqnarray}
\frac{Z}{Z_0}=\frac{Z_A}{\int [d\eta(r)]\,[d\eta(b)]\,e^{S_{0}(b,r)}}\,,
\label{ZA0}
\end{eqnarray}
with $Z_A$ defined by
\begin{eqnarray}
&&Z_A=\int [d\eta(r)]\,[d\eta(b)]\nonumber\\
&&\exp{\left(S_{0}(b,r)+N\,tr\ln\biggl(1-N^{-1}L_*^{-1}
\Bigl(g_1\,b+ g_2\,b^*-\sqrt{\lambda}\,r^*\Bigr)L^{-1}\Bigl(g_1\,b^*+
g_2\,b-\sqrt{\lambda}\,r\Bigr)\biggr)\right)}\,.
\label{ZA1}
\end{eqnarray}
We have arrived at an effective action in terms of two bosonic modes interacting in a highly non trivial way. Since we are interested in knowing the asymptotic behavior of the quotient $\frac{Z}{Z_0}$ in the
thermodynamic limit, i. e., $N\rightarrow\infty$ we analyze the asymptotic behavior of the last
defined expression $Z_A$. First, to make the dependence in $N$ more explicit, we rescale the bosonic
fields by $b\rightarrow\sqrt{N}\,b$, $b^*\rightarrow\sqrt{N}\,b^*$, $r\rightarrow\sqrt{N}\,r$
and $r^*\rightarrow\sqrt{N}\,r^*$. We get
\begin{eqnarray}
Z_A=A(N)\int [d\eta(r)]\,[d\eta(b)]\exp{\Bigl(N\,\Phi(b,b^*,r,r^*)\Bigr)}\,,
\label{ZA2}
\end{eqnarray}
with the function $\Phi(b,b^*,r,r^*)$ defined by
\begin{eqnarray}
\Phi(b,b^*,r,r^*)=S_{0}(b,r)+tr\ln\biggl(1-L_*^{-1}
\Bigl(g_1\,b+ g_2\,b^*-\sqrt{\lambda}\,r^*\Bigr)L^{-1}\Bigl(g_1\,b^*+ g_2\,b-\sqrt{\lambda}\,r\Bigr)\biggr)\,.
\label{fi1}
\end{eqnarray}
The term $A(N)$ in Eq. (\ref{ZA2}) comes from transforming the functional measures $[d\eta(r)]$ and $[d\eta(b)]$ under scaling
the bosonic field by $b\rightarrow\sqrt{N}\,b$, $b^*\rightarrow\sqrt{N}\,b^*$, $r\rightarrow\sqrt{N}\,r$
and $r^*\rightarrow\sqrt{N}\,r^*$ and will be reabsorbed when we transform back to the original variable.

The asymptotic behavior of the  functional integral appearing in Eq. (\ref{ZA2})
when $N\rightarrow\infty$, is obtained using the method of steepest descent \cite{amit}. We expand
the function $\Phi(b^*,b,r^*,r)$ around stationary points $b(\tau)=b_0(\tau)$,
$b^*(\tau)=b^*_0(\tau)$, $r(\tau)=r_0(\tau)$ and
$r^*(\tau)=r^*_0(\tau)$. 
The stationary points are solution of the equations 
\begin{eqnarray}\frac{\delta\,
\Phi(b,b^*,r,r^*)}{\delta\,b(\tau)}\Bigl|_{b_0,b^*_0,r_0,r^*_0}=0\ \ \ , 
\ \ \ \frac{\delta\,\Phi(b,b^*,r,r^*)}{\delta\,b^*(\tau)}\Bigl|_{b_0,b^*_0,r_0,r^*_0}=0\ \ \ ,\nonumber\\ \frac{\delta\,
\Phi(b,b^*,r,r^*)}{\delta\,r(\tau)}\Bigl|_{b_0,b^*_0,r_0,r^*_0}=0\ \ \  ,\ \ \  
\frac{\delta\,\Phi(b,b^*,r,r^*)}{\delta\,r^*(\tau)}\Bigl|_{b_0,b^*_0,r_0,r^*_0}=0\ \ \ .
\end{eqnarray}
which for the system at hand  are constant functions,  $b(\tau)=b_0$, $b^*(\tau)=b^*_0$,
$r(\tau)=r_0$ and $r^*(\tau)=r^*_0$ that fulfill 
the following system of algebraic equations,
\begin{eqnarray}
\omega_{0}\,b_0=
\biggl(g_1\Bigl(g_1\,b_0+ g_2\,b^*_0-\sqrt{\lambda}\,r^*_0\Bigr)+
g_2\Bigl(g_1\,b^*_0+ g_2\,b_0-\sqrt{\lambda}\,r_0\Bigr)\biggr)
\frac{1}{\Omega_{\Delta}}
\tanh{\left(\frac{\beta}{2}\Omega_{\Delta}\right)} \,,
\label{dphidbe2}
\end{eqnarray}
\begin{eqnarray}
\omega_{0}\,b^*_0=
\biggl(g_1\Bigl(g_1\,b^*_0+ g_2\,b_0-\sqrt{\lambda}\,r_0\Bigr)+
g_2\Bigl(g_1\,b_0+ g_2\,b^*_0-\sqrt{\lambda}\,r^*_0\Bigr)\biggr)
\frac{1}{\Omega_{\Delta}}
\tanh{\left(\frac{\beta}{2}\Omega_{\Delta}\right)} \,,
\label{dphidb2}
\end{eqnarray}
\begin{eqnarray}
r_0=\sqrt{\lambda}\,
\Bigl(g_1\,b^*_0+ g_2\,b_0-\sqrt{\lambda}\,r_0\Bigr)\,
\frac{1}{\Omega_{\Delta}}
\tanh{\left(\frac{\beta}{2}\Omega_{\Delta}\right)}  \,,
\label{dphidre2}
\end{eqnarray}
\begin{eqnarray}
r^*_0=\sqrt{\lambda}\,
\Bigl(g_1\,b_0+ g_2\,b^*_0-\sqrt{\lambda}\,r^*_0\Bigr)\,
\frac{1}{\Omega_{\Delta}}
\tanh{\left(\frac{\beta}{2}\Omega_{\Delta}\right)}  \,,
\label{dphidr2}
\end{eqnarray}
with $\Omega_{\Delta}=\sqrt{\,\Omega^2+4\,|g_1\,b_0+g_2\,b^*_0-\sqrt{\lambda}\,r^*_0\,|^2}$. 
Substituting Eqs. (\ref{dphidre2}) and (\ref{dphidr2}) in Eqs. (\ref{dphidbe2}) and (\ref{dphidb2})
we have also
\begin{eqnarray}
\sqrt{\lambda}\,\omega_{0}\,b_0&=&g_1r_0^*+g_2r_0 \,, \nonumber\\
\sqrt{\lambda}\,\omega_{0}\,b^*_0&=&g_1r_0+g_2r_0^* \, .
\label{dphidbe3}
\end{eqnarray}
Substituting  Eq. (\ref{dphidbe3}) in the Eq. (\ref{dphidbe2}) and imposing $b_0\neq 0$ we finally obtain that
\begin{eqnarray}
\frac{\omega_{0}}{(g_1+g_2)^2-\omega_0\,\lambda}= \frac{1}{\Omega_{\Delta}}
\tanh{\left(\frac{\beta}{2}\Omega_{\Delta}\right)}\,.
\label{eqfase}
\end{eqnarray}
This last equation allows us to calculate  $|b_0|$. Substituting Eq. (\ref{eqfase}) in Eq. (\ref{dphidbe2})
shows that $b_0$ is real and in consequence $r_0$ is real. The critical temperature
is characterized by the condition $b_0=0$. Therefore we have
\begin{eqnarray}
\frac{\omega_{0}\,\Omega}{(g_1+g_2)^2-\omega_0\,\lambda}=\tanh{\left(\frac{\beta_c}{2}\Omega\right)}\,.
\label{eqfase1}
\end{eqnarray}
At zero temperature this result reproduces the critical coupling of the quantum phase transition discussed in Ref. \cite{chen}. We stress that the introduction of the dipole-dipole interaction modifies the critical temperature of the transition from the fluorescent to the superradiant phase.

\section{The free energy}

Let us complete the computation of the asymptotic behavior of the functional integral, appearing in Eq. (\ref{ZA2}), in the thermodynamic limit, $N\rightarrow\infty$. We consider 
the two first leading terms in the  functional integral appearing in Eq. (\ref{ZA2}) coming from the expansion
of $\Phi(b,b^*,r,r^*)$ around the maximal value $b_0$, $b^*_0$, $r_0$ and $r^*_0$ this expansion is given by
\begin{eqnarray}
&&\Phi(b,b^*,r,r^*)=\Phi(b_0,b^*_0,r_0,r^*_0)+\frac{1}{2}\int_0^{\beta}d\tau_1\,d\tau_2\nonumber\\
&&\Bigl(b(\tau_1)-b_0\,,\,b^*(\tau_1)-b^*_0\,,\,r(\tau_1)-r_0\,,\,r^*(\tau_1)-r^*_0\Bigr)
\,\,M_{\Phi}\,\,
\left(\begin{array}{c}
b(\tau_2)-b_0\\
b^*(\tau_2)-b^*_0\\
r(\tau_2)-r_0\\
r^*(\tau_2)-r^*_0
\end{array}\right)\,,
\label{fi2}
\end{eqnarray}
the matrix $M_{\Phi}$, is given by
\begin{eqnarray}
\left.
M_{\Phi}=\left(\begin{array}{cccc}
\frac{\delta^2\Phi(b,b^*,r,r^*)}{\delta b(\tau_1)\,\delta b(\tau_2)}&
\frac{\delta^2\Phi(b,b^*,r,r^*)}{\delta b(\tau_1)\,\delta b^*(\tau_2)}&
\frac{\delta^2\Phi(b,b^*,r,r^*)}{\delta b(\tau_1)\,\delta r(\tau_2)}&
\frac{\delta^2\Phi(b,b^*,r,r^*)}{\delta b(\tau_1)\,\delta r^*(\tau_2)}
\\
\frac{\delta^2\Phi(b,b^*,r,r^*)}{\delta b^*(\tau_1)\,\delta b(\tau_2)}&
\frac{\delta^2\Phi(b,b^*,r,r^*)}{\delta b^*(\tau_1)\,\delta b^*(\tau_2)}&
\frac{\delta^2\Phi(b,b^*,r,r^*)}{\delta b^*(\tau_1)\,\delta r(\tau_2)}&
\frac{\delta^2\Phi(b,b^*,r,r^*)}{\delta b^*(\tau_1)\,\delta r^*(\tau_2)}
\\
\frac{\delta^2\Phi(b,b^*,r,r^*)}{\delta r(\tau_1)\,\delta b(\tau_2)}&
\frac{\delta^2\Phi(b,b^*,r,r^*)}{\delta r(\tau_1)\,\delta b^*(\tau_2)}&
\frac{\delta^2\Phi(b,b^*,r,r^*)}{\delta r(\tau_1)\,\delta r(\tau_2)}&
\frac{\delta^2\Phi(b,b^*,r,r^*)}{\delta r(\tau_1)\,\delta r^*(\tau_2)}
\\
\frac{\delta^2\Phi(b,b^*,r,r^*)}{\delta r^*(\tau_1)\,\delta b(\tau_2)}&
\frac{\delta^2\Phi(b,b^*,r,r^*)}{\delta r^*(\tau_1)\,\delta b^*(\tau_2)}&
\frac{\delta^2\Phi(b,b^*,r,r^*)}{\delta r^*(\tau_1)\,\delta r(\tau_2)}&
\frac{\delta^2\Phi(b,b^*,r,r^*)}{\delta r^*(\tau_1)\,\delta r^*(\tau_2)}
\end{array}\right)\,
\right|_{b_0,b^*_0,r_0,r^*_0}\,.
\label{Mfi}
\end{eqnarray}
Substituting this expansion given by Eq. (\ref{fi2}) in Eq. (\ref{ZA2}) we have that
\begin{eqnarray}
Z_A&=&e^{N\Phi(b^*_0,b_0)}\int[d\eta(r)]\,[d\eta(b)]\nonumber\\
&&\exp{\left(\frac{1}{2}\int_0^{\beta}d\tau_1\,d\tau_2\,
\Bigl(b(\tau_1)\,,\,b^*(\tau_1)\,,\,r(\tau_1)\,,\,r^*(\tau_1)\Bigr)\,M_{\Phi}
\left(\begin{array}{c}
b(\tau_2)\\
b^*(\tau_2)\\
r(\tau_2)\\
r^*(\tau_2)
\end{array}\right)\right)}\, \ \ .
\label{ZA3}
\end{eqnarray}
To obtain the last expression, we  applied the transformation $b(\tau)\rightarrow \Bigr(b(\tau)+b_0\Bigl)/\sqrt{N}$,  
$b^*(\tau)\rightarrow \Bigr(b^*(\tau)+b^*_0\Bigl)/\sqrt{N}$, $r(\tau)\rightarrow \Bigr(r(\tau)+r_0\Bigl)/\sqrt{N}$ and 
$r^*(\tau)\rightarrow \Bigr(r^*(\tau)+r^*_0\Bigl)/\sqrt{N}$ to the fields in the functional integral.
Terms of higher order in the fields in the expansion given by Eq. (\ref{fi2}) generates terms in Eq. (\ref{ZA3}) suppressed by powers of $1/\sqrt{N}$. Therefore in the thermodynamic limit the Eq. 
(\ref{fi2}) is a good approximation.

The free energy  $F=-\frac{1}{N\,\beta}\ln Z$,  in the
thermodynamic limit,  $N\rightarrow\infty$ is obtained from Eq. (\ref{ZA0})  using Eq. (\ref{ZA3}). In the normal phase $F-F_0=0$, where $F_0$ is the free energy for the non-interacting model. Finally, in the superradiant phase we have that
\begin{eqnarray}
F-F_0=\frac{\omega_0\,(\Omega_{\Delta}^2-\Omega^2)}{4\,\Bigl((g_1+g_2)^2-\lambda\,\omega_0\Bigr)}-\frac{1}{\beta}\,
\ln{\left(\frac{\cosh\left(\frac{\beta\,\Omega_{\Delta}}{2}\right)}
{\cosh\left(\frac{\beta\,\Omega}{2}\right)}\right)}\,.
\label{freenergySup}
\end{eqnarray}

\section{Conclusions}
 
Functional methods provide an efficient tool to compute the thermodynamics functions and critical properties of spin models. In the limit of zero temperature the results obtained by this means may be compared with the ones obtained directly from an exact or approximate quantum mechanical solution of the system involved.

In this  paper we compute the free energy of the full Dicke model with dipole-dipole interaction at temperature $\beta^{-1}$ and determine the critical temperature for the superradiant phase transition. To handle the non-Gaussian terms resulting from the presence of the dipole-dipole interaction we introduce a single auxiliary field which allow to compute an effective action in terms of two bosonic fields. This action showed to be manageable using the saddle point approximation and in this way we determine the corrections to the critical temperature of the full Dicke model computed in \cite{hioe,aparicio2}. This approach improves the approximation scheme  of Ref. \cite{nami} where no correction was detected. For the  full Dicke model with dipole-dipole interaction  addressed in this paper the critical coupling associated with the quantum superradiant phase transition were obtained in Refs. \cite{chen,nie} following \cite{emary} by performing a Holstein Primakoff and then the diagonalization of the effective Hamiltonian so obtained. At zero temperature our result reproduces the critical coupling of the quantum phase transition discussed in Ref.  \cite{chen,nie}.

In Refs. \cite{emary,chen,nie} the relation of the entanglement of the collective modes of the Dicke models  and the quantum phase transition was discussed. Since for finite temperature the thermal interaction usually induces decoherence in quantum systems it could be interesting to link this analysis with the thermodynamic behavior of the system. To this end, the excitation spectrum of the system should be computed and understood. This subject, as well as the effect of introducing disorder \cite{dis} in the dipole-dipole coupling are being investigated by the authors.

\section{Acknowledgments}

One of the authors (MAA) would like to thank the Institut f\"ur Theoretische Physik
of the Technische Universit\"at Berlin for their kind hospitality.
MAA acknowledges FAPESP, NFS acknowledges CNPq and JS acknowledges 
Project Did-Gid30 for financial support.

\end{document}